\newcommand{\CLASSINPUTtoptextmargin}{0.75in}
\newcommand{\CLASSINPUTbottomtextmargin}{1.0in}
\documentclass[conference]{IEEEtran}
\IEEEoverridecommandlockouts

\UseRawInputEncoding
\usepackage[top=0.75in,bottom=1.0in,left=0.68in,right=0.68in]{geometry}
\def\BibTeX{{\rm B\kern-.05em{\sc i\kern-.025em b}\kern-.08em T\kern-.1667em\lower.7ex\hbox{E}\kern-.125emX}}
\usepackage{cite}
\usepackage{amsmath,amssymb,amsfonts}
\usepackage{algorithmic}
\usepackage{graphicx}
\usepackage{textcomp}
\usepackage{xcolor}
\usepackage{amsthm}
\usepackage{amsmath}
\usepackage{amssymb}
\usepackage{booktabs}
\usepackage{amsfonts}
\usepackage{lipsum}
\usepackage[inkscapeformat=png]{svg}
\usepackage{mathtools}
\usepackage{threeparttable}
\usepackage{tablefootnote}
\usepackage[hyphens]{url}
\usepackage[hidelinks]{hyperref}
\usepackage{algorithmic, algorithm}
\usepackage{tabularx}
\usepackage{subfig}
\usepackage{mwe}
\usepackage{ragged2e}
\usepackage{dblfloatfix}
\usepackage{footmisc}
\usepackage{footnote}
\usepackage[labelfont=bf,font=small]{caption}
\makesavenoteenv{tabular}
\makesavenoteenv{table}

\usepackage[inline]{enumitem}

\makeatletter
\newcommand\fs@betterruled{%
  \def\@fs@cfont{\bfseries}\let\@fs@capt\floatc@ruled
  \def\@fs@pre{\vspace*{5pt}\hrule height.8pt depth0pt \kern2pt}%
  \def\@fs@post{\kern2pt\hrule\relax}%
  \def\@fs@mid{\kern2pt\hrule\kern2pt}%
  \let\@fs@iftopcapt\iftrue}
\floatstyle{betterruled}
\restylefloat{algorithm}
\makeatother

\definecolor{custom_blue}{HTML}{1F77B4}
\definecolor{custom_orange}{HTML}{FF7F0E}
\definecolor{custom_green}{HTML}{2CA02C}

\usepackage{tikz}
\usepackage{pgfplots}
\pgfplotsset{compat=1.17}

\newcommand{\gf}[1]{\textcolor{cyan}{{#1}}}

\newcommand{\mx}[1]{\mathbf{#1}}

\usepackage[nolist,printonlyused]{acronym} 
\renewcommand{\baselinestretch}{0.92}
\begin{document}

\begin{acronym}
  \acro{2G}{Second Generation}
  \acro{3G}{3$^\text{rd}$~Generation}
  \acro{3GPP}{3$^\text{rd}$~Generation Partnership Project}
  \acro{4G}{4$^\text{th}$~Generation}
  \acro{5G}{5$^\text{th}$~Generation}
  \acro{AA}{Antenna Array}
  \acro{AC}{Admission Control}
  \acro{AD}{Attack-Decay}
  \acro{ADSL}{Asymmetric Digital Subscriber Line}
	\acro{AHW}{Alternate Hop-and-Wait}
  \acro{AMC}{Adaptive Modulation and Coding}
	\acro{AP}{access point}
  \acro{AoI}{age of information}
  \acro{APA}{Adaptive Power Allocation}
  \acro{AR}{autoregressive}
  \acro{ARMA}{Autoregressive Moving Average}
  \acro{ATES}{Adaptive Throughput-based Efficiency-Satisfaction Trade-Off}
  \acro{AWGN}{additive white Gaussian noise}
  \acro{BB}{Branch and Bound}
  \acro{BD}{Block Diagonalization}
  \acro{BER}{bit error rate}
  \acro{BF}{Best Fit}
  \acro{BLER}{BLock Error Rate}
  \acro{BPC}{Binary power control}
  \acro{BPSK}{binary phase-shift keying}
  \acro{BPA}{Best \ac{PDPR} Algorithm}
  \acro{BRA}{Balanced Random Allocation}
  \acro{BS}{base station}
  \acro{CAP}{Combinatorial Allocation Problem}
  \acro{CAPEX}{Capital Expenditure}
  \acro{CAZAC}{Constant Amplitude Zero Autocorrelation}
  \acro{CB}{codebook}
  \acro{CBF}{Coordinated Beamforming}
  \acro{CBR}{Constant Bit Rate}
  \acro{CBS}{Class Based Scheduling}
  \acro{CC}{Congestion Control}
  \acro{CDF}{Cumulative Distribution Function}
  \acro{CDMA}{Code-Division Multiple Access}
  \acro{CL}{Closed Loop}
  \acro{CLPC}{Closed Loop Power Control}
  \acro{CNR}{Channel-to-Noise Ratio}
  \acro{CPA}{Cellular Protection Algorithm}
  \acro{CPICH}{Common Pilot Channel}
  \acro{CoMP}{Coordinated Multi-Point}
  \acro{CQI}{Channel Quality Indicator}
  \acro{CRLB}{Cram\'er-Rao Lower Bound}
  \acro{CRM}{Constrained Rate Maximization}
	\acro{CRN}{Cognitive Radio Network}
  \acro{CS}{Coordinated Scheduling}
  \acro{CSI}{channel state information}
  \acro{CSIR}{channel state information at the receiver}
  \acro{CSIT}{channel state information at the transmitter}
  \acro{CUE}{cellular user equipment}
  \acro{D2D}{device-to-device}
  \acro{DCA}{Dynamic Channel Allocation}
  \acro{DE}{Differential Evolution}
  \acro{DFT}{Discrete Fourier Transform}
  \acro{DISCOVER}{Deep Intrinsically Motivated Exploration}
  \acro{DIST}{Distance}
  \acro{DL}{downlink}
  \acro{DMA}{Double Moving Average}
	\acro{DMRS}{Demodulation Reference Signal}
  \acro{D2DM}{D2D Mode}
  \acro{DMS}{D2D Mode Selection}
  \acro{DPC}{Dirty Paper Coding}
  \acro{DRA}{Dynamic Resource Assignment}
  \acro{DRL}{deep reinforcement learning}
  \acro{DDPG}{Deep Deterministic Policy Gradient}
  \acro{RL}{reinforcement learning}
  \acro{SAC}{Soft Actor-Critic}
  \acro{DSA}{Dynamic Spectrum Access}
  \acro{DSM}{Delay-based Satisfaction Maximization}
  \acro{ECC}{Electronic Communications Committee}
  \acro{ECRB}{expectation of the conditional Cramér-Rao bound}
  \acro{EFLC}{Error Feedback Based Load Control}
  \acro{EI}{Efficiency Indicator}
  \acro{eNB}{Evolved Node B}
  \acro{EPA}{Equal Power Allocation}
  \acro{EPC}{Evolved Packet Core}
  \acro{EPS}{Evolved Packet System}
  \acro{E-UTRAN}{Evolved Universal Terrestrial Radio Access Network}
  \acro{ES}{Exhaustive Search}
  \acro{FDD}{frequency division duplexing}
  \acro{FDM}{Frequency Division Multiplexing}
  \acro{FER}{Frame Erasure Rate}
  \acro{FF}{Fast Fading}
  \acro{FIM}{Fisher Information Matrix}
  \acro{FSB}{Fixed Switched Beamforming}
  \acro{FST}{Fixed SNR Target}
  \acro{FTP}{File Transfer Protocol}
  \acro{GA}{Genetic Algorithm}
  \acro{GBR}{Guaranteed Bit Rate}
  \acro{GLR}{Gain to Leakage Ratio}
  \acro{GOS}{Generated Orthogonal Sequence}
  \acro{GPL}{GNU General Public License}
  \acro{GRP}{Grouping}
  \acro{HARQ}{Hybrid Automatic Repeat Request}
  \acro{HMS}{Harmonic Mode Selection}
  \acro{HOL}{Head Of Line}
  \acro{HSDPA}{High-Speed Downlink Packet Access}
  \acro{HSPA}{High Speed Packet Access}
  \acro{HTTP}{HyperText Transfer Protocol}
  \acro{ICMP}{Internet Control Message Protocol}
  \acro{ICI}{Intercell Interference}
  \acro{ID}{Identification}
  \acro{IETF}{Internet Engineering Task Force}
  \acro{ILP}{Integer Linear Program}
  \acro{JRAPAP}{Joint RB Assignment and Power Allocation Problem}
  \acro{UID}{Unique Identification}
  \acro{i.i.d.}{independent and identically distributed}
  \acro{IIR}{Infinite Impulse Response}
  \acro{ILP}{Integer Linear Problem}
  \acro{IMT}{International Mobile Telecommunications}
  \acro{INV}{Inverted Norm-based Grouping}
	\acro{IoT}{Internet of Things}
  \acro{IP}{Internet Protocol}
  \acro{IPv6}{Internet Protocol Version 6}
  \acro{IRS}{intelligent reflecting surface}
  \acro{ISD}{Inter-Site Distance}
  \acro{ISI}{Inter Symbol Interference}
  \acro{ITU}{International Telecommunication Union}
  \acro{JOAS}{Joint Opportunistic Assignment and Scheduling}
  \acro{JOS}{Joint Opportunistic Scheduling}
  \acro{JP}{Joint Processing}
	\acro{JS}{Jump-Stay}
    \acro{KF}{Kalman filter}
  \acro{KKT}{Karush-Kuhn-Tucker}
  \acro{L3}{Layer-3}
  \acro{LAC}{Link Admission Control}
  \acro{LA}{Link Adaptation}
  \acro{LC}{Load Control}
  \acro{LMMSE}{linear minimum mean squared error}
  \acro{LOS}{line of sight}
  \acro{LP}{Linear Programming}
  \acro{LS}{least squares}
  \acro{LTE}{Long Term Evolution}
  \acro{LTE-A}{LTE-Advanced}
  \acro{LTE-Advanced}{Long Term Evolution Advanced}
  \acro{M2M}{Machine-to-Machine}
  \acro{MAC}{Medium Access Control}
  \acro{MANET}{Mobile Ad hoc Network}
  \acro{MC}{Modular Clock}
  \acro{MCS}{Modulation and Coding Scheme}
  \acro{MDB}{Measured Delay Based}
  \acro{MDI}{Minimum D2D Interference}
  \acro{MF}{Matched Filter}
  \acro{MG}{Maximum Gain}
  \acro{MH}{Multi-Hop}
  \acro{MIMO}{multiple input multiple output}
  \acro{MINLP}{Mixed Integer Nonlinear Programming}
  \acro{MIP}{Mixed Integer Programming}
  \acro{MISO}{Multiple Input Single Output}
  \acro{ML}{maximum likelihood}
  \acro{MLWDF}{Modified Largest Weighted Delay First}
  \acro{MME}{Mobility Management Entity}
  \acro{MML}{misspecified maximum likelihood}
  \acro{MMSE}{minimum mean squared error}
  \acro{MOS}{Mean Opinion Score}
  \acro{MPF}{Multicarrier Proportional Fair}
  \acro{MRA}{Maximum Rate Allocation}
  \acro{MR}{Maximum Rate}
  \acro{MRC}{maximum ratio combining}
  \acro{MRT}{Maximum Ratio Transmission}
  \acro{MRUS}{Maximum Rate with User Satisfaction}
  \acro{MS}{mobile station}
  \acro{MSE}{mean squared error}
  \acro{MSI}{Multi-Stream Interference}
  \acro{MTC}{Machine-Type Communication}
  \acro{MTSI}{Multimedia Telephony Services over IMS}
  \acro{MTSM}{Modified Throughput-based Satisfaction Maximization}
  \acro{MU-MIMO}{multiuser multiple input multiple output}
  \acro{MU-MISO}{multiuser multiple input single output}
  \acro{MU}{multi-user}
  \acro{NAS}{Non-Access Stratum}
  \acro{NB}{Node B}
  \acro{NE}{Nash equilibrium}
  \acro{NCL}{Neighbor Cell List}
  \acro{NLP}{Nonlinear Programming}
  \acro{NLOS}{Non-Line of Sight}
  \acro{NMSE}{normalized mean squared error}
  \acro{NOMA}{non-orthogonal multiple access}
  \acro{NORM}{Normalized Projection-based Grouping}
  \acro{NP}{Non-Polynomial Time}
  \acro{NR}{New Radio}
  \acro{NRT}{Non-Real Time}
  \acro{NSPS}{National Security and Public Safety Services}
  \acro{O2I}{Outdoor to Indoor}
  \acro{OFDMA}{orthogonal frequency division multiple access}
  \acro{OFDM}{orthogonal frequency division multiplexing}
  \acro{OFPC}{Open Loop with Fractional Path Loss Compensation}
	\acro{O2I}{Outdoor-to-Indoor}
  \acro{OL}{Open Loop}
  \acro{OLPC}{Open-Loop Power Control}
  \acro{OL-PC}{Open-Loop Power Control}
  \acro{OPEX}{Operational Expenditure}
  \acro{ORB}{Orthogonal Random Beamforming}
  \acro{JO-PF}{Joint Opportunistic Proportional Fair}
  \acro{OSI}{Open Systems Interconnection}
  \acro{PAIR}{D2D Pair Gain-based Grouping}
  \acro{PAPR}{Peak-to-Average Power Ratio}
  \acro{P2P}{Peer-to-Peer}
  \acro{PC}{Power Control}
  \acro{PCI}{Physical Cell ID}
  \acro{PDF}{Probability Density Function}
  \acro{PDPR}{pilot-to-data power ratio}
  \acro{PER}{Packet Error Rate}
  \acro{PF}{Proportional Fair}
  \acro{P-GW}{Packet Data Network Gateway}
  \acro{PL}{Pathloss}
  \acro{PPR}{pilot power ratio}
  \acro{PRB}{physical resource block}
  \acro{PROJ}{Projection-based Grouping}
  \acro{ProSe}{Proximity Services}
  \acro{PS}{Packet Scheduling}
  \acro{PSAM}{pilot symbol assisted modulation}
  \acro{PSO}{Particle Swarm Optimization}
  \acro{PZF}{Projected Zero-Forcing}
  \acro{QAM}{Quadrature Amplitude Modulation}
  \acro{QoS}{Quality of Service}
  \acro{QPSK}{Quadri-Phase Shift Keying}
  \acro{RAISES}{Reallocation-based Assignment for Improved Spectral Efficiency and Satisfaction}
  \acro{RAN}{Radio Access Network}
  \acro{RA}{Resource Allocation}
  \acro{RAT}{Radio Access Technology}
  \acro{RATE}{Rate-based}
  \acro{RB}{resource block}
  \acro{RBG}{Resource Block Group}
  \acro{REF}{Reference Grouping}
  \acro{RIS}{reconfigurable intelligent surface}
  \acro{RLC}{Radio Link Control}
  \acro{RM}{Rate Maximization}
  \acro{RNC}{Radio Network Controller}
  \acro{RND}{Random Grouping}
  \acro{RRA}{Radio Resource Allocation}
  \acro{RRM}{Radio Resource Management}
  \acro{RSCP}{Received Signal Code Power}
  \acro{RSRP}{Reference Signal Receive Power}
  \acro{RSRQ}{Reference Signal Receive Quality}
  \acro{RR}{Round Robin}
  \acro{RRC}{Radio Resource Control}
  \acro{RSSI}{Received Signal Strength Indicator}
  \acro{RT}{Real Time}
  \acro{RU}{Resource Unit}
  \acro{RUNE}{RUdimentary Network Emulator}
  \acro{RV}{Random Variable}
  \acro{SAC}{Soft Actor-Critic}
  \acro{SCM}{Spatial Channel Model}
  \acro{SC-FDMA}{Single Carrier - Frequency Division Multiple Access}
  \acro{SD}{Soft Dropping}
  \acro{S-D}{Source-Destination}
  \acro{SDPC}{Soft Dropping Power Control}
  \acro{SDMA}{Space-Division Multiple Access}
  \acro{SE}{spectral efficiency}
  \acro{SER}{Symbol Error Rate}
  \acro{SES}{Simple Exponential Smoothing}
  \acro{S-GW}{Serving Gateway}
  \acro{SINR}{signal-to-interference-plus-noise ratio}
  \acro{SI}{Satisfaction Indicator}
  \acro{SIP}{Session Initiation Protocol}
  \acro{SISO}{single input single output}
  \acro{SIMO}{single input multiple output}
  \acro{SIR}{signal-to-interference ratio}
  \acro{SLNR}{Signal-to-Leakage-plus-Noise Ratio}
  \acro{SMA}{Simple Moving Average}
  \acro{SNR}{signal-to-noise ratio}
  \acro{SORA}{Satisfaction Oriented Resource Allocation}
  \acro{SORA-NRT}{Satisfaction-Oriented Resource Allocation for Non-Real Time Services}
  \acro{SORA-RT}{Satisfaction-Oriented Resource Allocation for Real Time Services}
  \acro{SPF}{Single-Carrier Proportional Fair}
  \acro{SRA}{Sequential Removal Algorithm}
  \acro{SRS}{Sounding Reference Signal}
  \acro{SU-MIMO}{single-user multiple input multiple output}
  \acro{SU}{Single-User}
  \acro{SVD}{Singular Value Decomposition}
  \acro{TCP}{Transmission Control Protocol}
  \acro{TDD}{time division duplexing}
  \acro{TDMA}{Time Division Multiple Access}
  \acro{TETRA}{Terrestrial Trunked Radio}
  \acro{TP}{Transmit Power}
  \acro{TPC}{Transmit Power Control}
  \acro{TTI}{Transmission Time Interval}
  \acro{TTR}{Time-To-Rendezvous}
  \acro{TSM}{Throughput-based Satisfaction Maximization}
  \acro{TU}{Typical Urban}
  \acro{UAV}{unmanned aerial vehicle}
  \acro{UE}{user equipment}
  \acro{UEPS}{Urgency and Efficiency-based Packet Scheduling}
  \acro{UL}{uplink}
  \acro{UMTS}{Universal Mobile Telecommunications System}
  \acro{URI}{Uniform Resource Identifier}
  \acro{URM}{Unconstrained Rate Maximization}
  \acro{UT}{user terminal}
  \acro{VAR}{vector autoregressive}
  \acro{VR}{Virtual Resource}
  \acro{VoIP}{Voice over IP}
  \acro{WAN}{Wireless Access Network}
  \acro{WCDMA}{Wideband Code Division Multiple Access}
  \acro{WF}{Water-filling}
  \acro{WiMAX}{Worldwide Interoperability for Microwave Access}
  \acro{WINNER}{Wireless World Initiative New Radio}
  \acro{WLAN}{Wireless Local Area Network}
  \acro{WMPF}{Weighted Multicarrier Proportional Fair}
  \acro{WPF}{Weighted Proportional Fair}
  \acro{WSN}{Wireless Sensor Network}
  \acro{WSS}{wide-sense stationary}
  \acro{WWW}{World Wide Web}
  \acro{XIXO}{(Single or Multiple) Input (Single or Multiple) Output}
  \acro{ZF}{zero-forcing}
  \acro{ZMCSCG}{Zero Mean Circularly Symmetric Complex Gaussian}
\end{acronym}

\title{Impact of Pilot Contamination Between Operators With Interfering Reconfigurable Intelligent Surfaces}
\author{\IEEEauthorblockN{Do\u{g}a G\"{u}rg\"{u}no\u{g}lu\IEEEauthorrefmark{1}, Emil Bj\"{o}rnson\IEEEauthorrefmark{1},  G\'{a}bor Fodor\IEEEauthorrefmark{1}\IEEEauthorrefmark{2}}
\IEEEauthorblockA{\IEEEauthorrefmark{1}KTH Royal Institute of Technology, Stockholm 100 44, Sweden\\\IEEEauthorrefmark{2}Ericsson Research, Stockholm 164 80, Sweden}
\thanks{This study is supported by EU Horizon 2020 MSCA-ITN-METAWIRELESS, Grant Agreement 956256. E.~Bj\"ornson was supported by FFL18-0277 grant from the Swedish Foundation for Strategic Research.}
\vspace{-0.1cm}}

\maketitle

\begin{abstract}
In this paper, we study the impact of pilot contamination in a system 
where two operators serve their respective users with the assistance of two wide-band reconfigurable intelligent surfaces (RIS), each belonging to a single operator. We consider one active user per operator and they use disjoint narrow frequency bands. Although each RIS is dedicated to a single operator, both users' transmissions are reflected by both RISs. We show that this creates a new kind of pilot contamination effect when pilots are transmitted simultaneously. Since combating inter-operator pilot contamination in RIS-assisted networks would require long pilot signal sequences to maintain orthogonality among the users of different operators, we propose the orthogonal configurations of the RISs. Numerical results show that this approach completely eliminates pilot contamination, and significantly improves the performance in terms of channel estimation and equalization by removing the channel estimation bias.
\end{abstract}


\section{Introduction}
Pilot contamination is a key problem that frequently arises in wireless communication systems \cite{Marzetta2010a}. When multiple users use the same pilot sequences simultaneously in the same band, due to the limited channel coherence time, the \ac{BS} cannot distinguish their channels, which typically results in poor channel estimates and extra beamformed interference towards pilot-sharing \acp{UE}. Therefore, pilot contamination adversely affects the coherent reception of data, and methods to mitigate pilot contamination have been widely studied in the communication literature \cite{Marzetta2010a,Sanguinetti2020a,pCon1,Saxena:15}.

In recent years, \acp{RIS} have arisen as a new technology component for 6G \cite{ris_commag}. An \ac{RIS} is a surface consisting of multiple reflecting elements that have sub-wavelength spacing and controllable  reflection properties \cite{Bjornson2020a}.
This feature provides partial control of
the propagation environment that can lead to better services for users, especially when their serving \ac{BS} is not in their \ac{LOS}. By adjusting the impedances of the individual elements via a \ac{RIS} controller, the elements are capable of adding desired phase shifts to the reflected signals, thereby forming reflected beams in desired directions that can significantly boost the \ac{SINR} at the receiver \cite{ris_commag}.

On the other hand, the addition of \acp{RIS} to existing systems introduces new design and operational challenges. For example, the length of the pilot signal required by a single \ac{UE} is proportional to the number of \ac{RIS} elements (e.g., tens or hundreds), because the \ac{RIS} must change its configuration to explore all channel dimensions \cite{risChanEst,Bjornson2022b}. 
In addition, passive \ac{RIS} causes multiplicative path losses, which increases the large-scale fading loss between the transmitter and the receiver \cite{ris_zappone}. Active \acp{RIS} \cite{georgios}, on the other hand, are less energy efficient, and due to the presence of amplifiers, it introduces additional noise. While the aforementioned problems caused by the \ac{RIS} have been recognized \cite{Garg:22}, pilot contamination caused by the presence of multiple \acp{RIS} has not been studied in the literature. In this paper, however, we identify the creation of pilot contamination as another practical challenge: a \ac{UE} that transmits pilots to its serving \ac{BS} via multiple \ac{RIS}, which change their configurations simultaneously, may cause a new kind of pilot contamination that has not been studied. 

Wireless communication systems are governed by standards, and physical-layer specifications usually contain pre-defined sequences for pilot signals and codebooks for directional beamforming \cite{38211}. While the use \acp{RIS} is not standardized yet, it is likely that the configuration sequences that facilitate the deployment of \acp{RIS} while maintaining interoperability will be standardized. Consequently, when multiple cellular networks are deployed by different network operators in overlapping geographical areas, the \acp{RIS} may adopt identical or overlapping pilot sequences and cause pilot contamination. 

Due to the ability to change the environment's propagation characteristics, deploying multiple \acp{RIS} in a  geographical area also implies that pilot contamination can occur due to a \ac{UE}'s own pilot signal. 
Since this phenomenon may exacerbate the pilot contamination problem, it is clear that pilot contamination due to the presence of
multiple \acp{RIS} must be dealt with.

In this paper, we study the pilot contamination caused by the presence of multiple \acp{RIS} by considering the uplink of a system consisting of two wide-band \acp{RIS}, two single-antenna \acp{UE}, and two co-located single-antenna \acp{BS}, where the two \acp{UE} are subscribed to different operators with non-overlapping narrow-band channels at different frequencies. Each \ac{RIS} is dedicated to a single operator, but both \ac{UE} signals are reflected from both \acp{RIS}. In this scenario, although there is no interference between the two \acp{UE}, both \acp{RIS} affect both frequency bands. We propose the use of orthogonal \ac{RIS} configuration sequences during pilot transmission to avoid pilot contamination. First, we describe the channel estimation procedure with a lack of information on channel statistics, i.e., when the channels are characterized  by deterministic parameters. Assuming identical and orthogonal \ac{RIS} configurations, we derive the statistics of the \ac{ML} estimates and the effect of pilot contamination on channel estimation. 
Then we analyze how it affects the data estimation process, and discuss how this effect can be mitigated.

\section{System Model}\label{sec:system_model}
\begin{figure}
    \centering
    \includegraphics[width=0.84\linewidth]{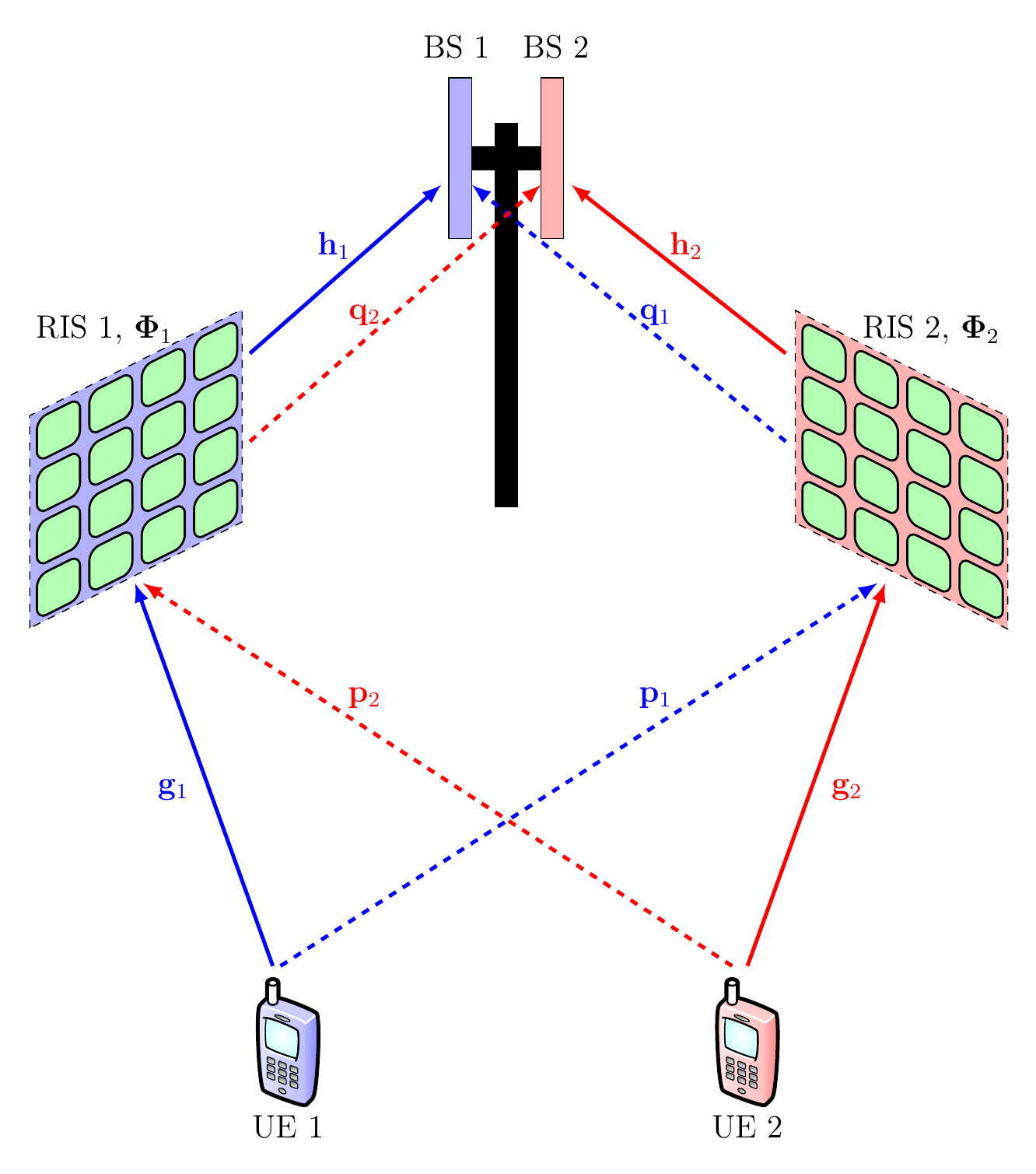}
    \caption{System setup with two \ac{UE}s, two \acp{RIS} and two co-located single-antenna \acp{BS}. The blue channels correspond to frequency band 1, and the red channels correspond to frequency band 2, subscribed by \ac{UE}s $1$ and $2$, respectively. Desired channels are denoted by solid lines, while the undesired channels whose existence are unknown to the \acp{BS} are denoted by dashed lines. Each channel vector is $N$-dimensional in line with the number of elements in each \ac{RIS}.}
    \label{fig:systemSetup}
\end{figure}

We consider the uplink of a cellular system consisting of two wide-band \ac{RIS}, two single-antenna \ac{UE}, and two co-located single-antenna \acp{BS}, as shown in Fig.~\ref{fig:systemSetup}, where each \ac{RIS} has $N$ reflecting elements. The two \acp{UE} are subscribed to different operators---who use site-sharing to reduce deployment costs---and transmit over two disjoint narrow frequency bands to their respective serving \acp{BS}. Each \ac{RIS} is dedicated to and controlled by a single operator but affects both bands. We consider an environment where the direct \ac{UE}-\ac{BS} paths are blocked, while the \ac{UE}-\ac{RIS} and \ac{RIS}-\ac{BS} paths are operational. Since the \acp{BS} and \acp{RIS} have fixed deployment locations, we assume the \ac{RIS}-\ac{BS} channels $\mathbf{h}_k$ are known, while the \ac{UE}-\ac{RIS} channels $\mathbf{g}_k$ are unknown and to be estimated for $k = 1,2$.

The signal transmitted by \ac{UE} $k$ reaches its serving \ac{BS} through  the channels $\mathbf{h}_k$ and $\mathbf{g}_k$, for $k = 1,2$. Importantly, the \acp{UE}' transmitted signals are also reflected by the non-serving operator's \acp{RIS} towards their serving \acp{BS}, which contaminates the pilot signal reflected by the serving \ac{RIS}. This phenomenon is illustrated in Fig.~\ref{fig:systemSetup}, where the resulting \ac{UE}-\ac{RIS} and \ac{RIS}-\ac{BS} channels are denoted by $\mathbf{p}_k$ and $\mathbf{q}_k$, respectively, for $k = 1,2$.

Since the \acp{BS} are unaware of the channels $\mathbf{q}_k$ and $\mathbf{p}_k$, they adopt misspecified system models for the received pilot and data signals. We also assume that the prior distributions of the channels are unavailable. Consequently, the \acp{BS} estimate the channels $\mathbf{g}_1$ and $\mathbf{g}_2$ via classical non-Bayesian parameter estimation methods during the channel estimation phase and use the channel estimates to perform data estimation \cite[Section IV.C]{poorbook} during the data transmission phase.
Defining the pilot signal of \ac{UE} $k$ as $s_k\in\mathbb{C}$, the received pilot signals on bands $1$ and $2$ at the \acp{BS} can be expressed as
\begin{subequations}\label{eq:rxPilot}
\begin{align}
    &y_{p1} = \sqrt{P_p}\mathbf{h}_1^T\boldsymbol{\Phi}_1\mathbf{g}_1s_1+\sqrt{P_p}\mathbf{q}_1^T\boldsymbol{\Phi}_2\mathbf{p}_1s_1+w_{p1},\\
    &y_{p2} = \sqrt{P_p}\mathbf{h}_2^T\boldsymbol{\Phi}_2\mathbf{g}_2s_2+\sqrt{P_p}\mathbf{q}_2^T\boldsymbol{\Phi}_1\mathbf{p}_2s_2+w_{p2},
\end{align}
\end{subequations}
where $y_{pk}$ denotes the received pilot signal, $w_{pk}\sim\mathcal{CN}(0,1)$ denotes the receiver noise for band $k$, and $\boldsymbol{\Phi}_k = \text{diag}(e^{-j\phi_{k1}},\dots,e^{-j\phi_{kN}})$ denotes the $k$th \ac{RIS}'s response matrix. We assume $s_1=s_2=1$ without loss of generality. For channel estimation, it is more convenient to rewrite \eqref{eq:rxPilot} as
\begin{subequations}\label{eq:rxPilot2}
\begin{align}
    &y_{p1} = \sqrt{P_p}\boldsymbol{\phi}_1^T\mathbf{D}_{\mathbf{h}_1}\mathbf{g}_1+\sqrt{P_p}\boldsymbol{\phi}_2^T\mathbf{D}_{\mathbf{q}_1}\mathbf{p}_1+w_{p1},\\
    &y_{p2} = \sqrt{P_p}\boldsymbol{\phi}_2^T\mathbf{D}_{\mathbf{h}_2}\mathbf{g}_2+\sqrt{P_p}\boldsymbol{\phi}_1^T\mathbf{D}_{\mathbf{q}_2}\mathbf{p}_2+w_{p2},
\end{align}
\end{subequations}
where $\mathbf{D}_{\mathbf{h}_k}$ and $\mathbf{D}_{\mathbf{q}_k}$ represent the diagonal matrices containing the elements of $\mathbf{h}_k$ and $\mathbf{q}_k$, and $\boldsymbol{\phi}_k$ denotes the column vectors containing the diagonal entries of $\boldsymbol{\Phi}_k$ for $ k = 1,2$.

As there are $N$ parameters in $\mathbf{g}_1$ and $\mathbf{g}_2$, at least $N$ linearly independent observations are needed to estimate them. To this end, we perform $L\geq N$ pilot transmissions over time, and we vertically stack the received pilot signals to obtain
\begin{subequations}\label{eq:rxPilot}
\begin{align}
    &\mathbf{y}_{p1}=\sqrt{P_p}\mathbf{B}_1\mathbf{D}_{\mathbf{h}_1}\mathbf{g}_1+\sqrt{P_p}\mathbf{B}_2\mathbf{D}_{\mathbf{q}_1}\mathbf{p}_1+\mathbf{w}_{p1},\\
    &\mathbf{y}_{p2}=\sqrt{P_p}\mathbf{B}_2\mathbf{D}_{\mathbf{h}_2}\mathbf{g}_2+\sqrt{P_p}\mathbf{B}_1\mathbf{D}_{\mathbf{q}_2}\mathbf{p}_2+\mathbf{w}_{p2},
\end{align}
\end{subequations}
where $\mx{y}_{pk}=[y_{pk}[1],\dots,y_{pk}[L]]^T$ denotes the sequence of received uplink pilots from the $k$th \ac{UE} over $L$ time instances, and the matrices $\mx{B}_1$ and $\mx{B}_2$ represent the sequence of \ac{RIS} configurations over $L$ time instances, that is, $\mx{B}_k \triangleq \begin{bmatrix}\boldsymbol{\phi}_k[1] & \dots & \boldsymbol{\phi}_k[L]\end{bmatrix}^T$ for $k = 1,2$. Recall that \ac{BS} $1$ is not aware of the reflection of \ac{UE} $1$'s signal from \ac{RIS} $2$ and vice versa. Consequently, the \acp{BS} assumes the following misspecified received pilot signal models:
\begin{subequations}\label{eq:rxPilotHat}
\begin{align}
    &\hat{\mathbf{y}}_{p1}=\sqrt{P_p}\mathbf{B}_1\mathbf{D}_{\mathbf{h}_1}\mathbf{g}_1+\mathbf{w}_{p1},\\
    &\hat{\mathbf{y}}_{p2}=\sqrt{P_p}\mathbf{B}_2\mathbf{D}_{\mathbf{h}_2}\mathbf{g}_2+\mathbf{w}_{p2}\gf{.}
\end{align}
\end{subequations}
Based on \eqref{eq:rxPilotHat}, the estimation of $\mx{g}_1$ and $\mx{g}_2$ is described and analyzed in the next section.

\section{Maximum Likelihood Channel Estimation}
\label{sec:MLE}
To estimate $\mx{g}_k$, which is $N$-dimensional, \ac{BS} $k$ requires at least $N$ independent observations for $k = 1,2$. Hence, both $\mathbf{B}_1,\mathbf{B}_2\in\mathbb{C}^{L\times N}$ must have full column rank. Furthermore, we require that the \ac{RIS} configurations on different time instances be orthogonal and contain entries on the unit circle that can be realized using a reflecting element. These assumptions result in $\mx{B}_k^H\mx{B}_k=L\mx{I}_{N}$. In classical non-Bayesian parameter estimation, the \ac{ML} estimator is widely used, which maximizes the likelihood function of the received observation over the unknown parameter. Since the \acp{BS} have misspecified received pilot signal models, they will instead maximize the likelihood functions obtained from the misspecified model, leading to \ac{MML} estimators. For \eqref{eq:rxPilotHat}, the \ac{MML} estimator can be expressed as
\begin{align}\nonumber
    \hat{\mx{g}}_k &= \arg\max_{\mx{g}_k}f(\mx{y}_{pk};\mx{g}_k)\\
    &=\arg\max_{\mx{g}_k}\frac{1}{(\pi\sigma_w^2)^{L}}\exp{\left(-\frac{\|\mx{y}_{pk}-\sqrt{P_p}\mx{B}_k\mx{D}_{\mx{h}_k}\mx{g}_k\|^2}{\sigma_w^2}\right)}\nonumber\\
    &=\arg\min_{\mx{g}_k}\left\|\mx{y}_{pk}-\sqrt{P_p}\mx{B}_k\mx{D}_{\mx{h}_k}\mx{g}_k\right\|^2\nonumber\\
    &=\frac{1}{\sqrt{P_p}}\mathbf{D}_{\mathbf{h}_k}^{-1}(\mathbf{B}_k^H\mathbf{B}_k)^{-1}\mathbf{B}_k^H\mathbf{y}_{pk}\nonumber\\
    &=\frac{1}{L\sqrt{P_p}}\mathbf{D}_{\mathbf{h}_k}^{-1}\mathbf{B}_k^H\mathbf{y}_{pk}.\label{eq:gHatMLmismatch}
\end{align}
In the following subsections, we describe the behavior of this estimator for two different choices of the $\mx{B}_k$ matrices.

\subsection{Case 1: The \acp{RIS} Adopt the Same 
Configuration Sequence}
We discussed earlier that in the absence of inter-operator cooperation, it is highly likely that the \acp{RIS} will use the same sequence of configurations during the channel estimation phase, which corresponds to $\mx{B}_1=\mx{B}_2=\mx{B}$.\footnote{The analysis in this paper can be easily extended to the case when $\mx{B}_1= \mx{U}\mx{B}_2$ for some unitary matrix $\mx{U}$, so that configuration sequences have identical spans. It is the overlap of the spans that can cause issues.} In this case, \eqref{eq:gHatMLmismatch} becomes
\begin{align}
    \hat{\mx{g}}_k = \mx{g}_k + \mathbf{D}_{\mathbf{h}_k}^{-1}\mathbf{D}_{\mathbf{q}_k}\mathbf{p}_k+\frac{1}{L\sqrt{P}_p}\mathbf{D}_{\mx{h}_k}^{-1}\mathbf{B}^H\mathbf{w}_{pk}.
\end{align}
Since we consider the channels as deterministic parameters, we obtain the probability distribution 
\begin{equation}\label{eq:gHat_1}
    \hat{\mx{g}}_k\sim\mathcal{CN}\!\left(\mx{g}_k + \mathbf{D}_{\mathbf{h}_k}^{-1}\mathbf{D}_{\mathbf{q}_k}\mathbf{p}_k,\frac{\sigma_w^2}{LP_p}(\mathbf{D}_{\mathbf{h}_k}^H\mathbf{D}_{\mathbf{h}_k})^{-1}\!\right)
\end{equation}
We notice that $\hat{\mx{g}}_k$ is biased; that is, $\mx{b}_k\triangleq\mathbb{E}[\hat{\mx{g}}_k-\mx{g}_k]=\mathbf{D}_{\mathbf{h}_k}^{-1}\mathbf{D}_{\mathbf{q}_k}\mathbf{p}_k\neq\boldsymbol{0}$. The estimator bias does not vanish when increasing $P_p$ or $L$,  and decreasing $\sigma_w^2$, hence, it is not asymptotically unbiased. This is a new instance of an extensively studied phenomenon in the massive \ac{MIMO} literature: pilot contamination \cite{Marzetta2010a,Sanguinetti2020a}.
Interestingly, the \acp{RIS} cause pilot contamination even between two non-overlapping frequency bands, which has not been widely recognized in the literature so far.

\subsection{Case 2: 
The \acp{RIS} Adopt Different Configuration Sequences}
In this subsection, we consider the generic case of $\mx{B}_1\neq\mx{B}_2$. To motivate the proposed method for configuring $\mx{B}_1$ and $\mx{B}_2$, we first consider the case where the \acp{BS} are aware of the true signal model in \eqref{eq:rxPilot}, and therefore can estimate both $\mx{g}_k$ and $\mx{r}_k\triangleq\mx{D}_{\mx{q}_k}\mx{p_k}$. The resulting system model can be expressed as
\begin{subequations}\label{eq:fullCaseChanEst}
    \begin{align}
        &\mx{y}_{p1} = \sqrt{P}_p\begin{bmatrix}
            \mx{B}_1\mx{D}_{\mx{h}_1} & \mx{B}_2
        \end{bmatrix}\begin{bmatrix}
            \mx{g}_1\\\mx{r}_1
        \end{bmatrix} + \mx{w}_{p1},\\
        &\mx{y}_{p2} = \sqrt{P}_p\begin{bmatrix}
            \mx{B}_2\mx{D}_{\mx{h}_2} & \mx{B}_1
        \end{bmatrix}\begin{bmatrix}
            \mx{g}_2\\\mx{r}_2
        \end{bmatrix} + \mx{w}_{p2}.
    \end{align}    
\end{subequations}
In \eqref{eq:fullCaseChanEst}, a known linear transformation is applied to the parameter vector of interest in the presence of additive noise. Consequently, the \ac{ML} estimates for \ac{UE} 1's unknown channels become
\begin{equation}\label{eq:fullCaseML}
    \begin{bmatrix}
        \hat{\mx{g}}_1\\\hat{\mx{r}}_1
    \end{bmatrix} = \frac{1}{\sqrt{P_p}}\begin{bmatrix}
        L\mx{D}_{\mx{h}_1}^H\mx{D}_{\mx{h}_1} & \mx{D}_{\mx{h}_1}^H\mx{B}_1^H\mx{B}_2\\
        \mx{B}_2^H\mx{B}_1\mx{D}_{\mx{h}_1} & L\mx{I}_{N}
    \end{bmatrix}^{-1}\begin{bmatrix}
        \mx{D}_{\mx{h}_1}^H\mx{B}_1^H \\ \mx{B}_2^H
    \end{bmatrix}\mx{y}_{p1}.
\end{equation}
Note that in this case, the total dimension of the unknown parameter vector is $2N$, hence, at least $2N$ independent observations are required for the matrix inverse to exist. The structure in \eqref{eq:fullCaseML} applies to \ac{UE} 2 with alternated indices, and it gives the \ac{ML} estimator, which is both unbiased and efficient, since \eqref{eq:fullCaseChanEst} is a linear observation model with additive Gaussian noise \cite[Theorem 7.3]{kayestimation}. Hence, \eqref{eq:fullCaseML} is unbiased irrespective of other parameters such as $\sigma_w^2$, $L$, $P_p$, and it achieves the \ac{CRLB}, which provides a lower bound on the \ac{MSE} of any unbiased estimator \cite{poorbook}. It has to be noted  that when $\mx{B}_1^H\mx{B}_2 = \boldsymbol{0}$, \eqref{eq:fullCaseML} becomes
\begin{align}\nonumber\label{eq:fullCaseMLOrth}
    \begin{bmatrix}
        \hat{\mx{g}}_1\\\hat{\mx{r}}_1
    \end{bmatrix} &= \frac{1}{L\sqrt{P_p}}\begin{bmatrix}
        \mx{D}_{\mx{h}_1}^H\mx{D}_{\mx{h}_1} & \boldsymbol{0}\\
        \boldsymbol{0} & \mx{I}_{N}
    \end{bmatrix}^{-1}\begin{bmatrix}
        \mx{D}_{\mx{h}_1}^H\mx{B}_1^H \\ \mx{B}_2^H
    \end{bmatrix}\mx{y}_{p1}\\
    &= \frac{1}{L\sqrt{P_p}}\begin{bmatrix}
        \mx{D}_{\mx{h}_1}^{-1}\mx{B}_1^H \\ \mx{B}_2^H
    \end{bmatrix}\mx{y}_{p1}.
\end{align}
Note that the expression for $\hat{\mx{g}}_1$ in \eqref{eq:fullCaseMLOrth} is the same as in \eqref{eq:gHatMLmismatch}. This shows that when $\mx{B}_1^H\mx{B}_2 = \boldsymbol{0}$, the \ac{MML} in \eqref{eq:gHatMLmismatch} coincides with the \ac{ML} estimator; that is, the misspecified model is sufficient when the configuration sequences are designed to alleviate pilot interference. The probability distribution of $\hat{\mx{g}}_k$ in this case can be expressed as
\begin{equation}\label{eq:gHat_2}
    \hat{\mx{g}}_k\sim\mathcal{CN}\left(\mx{g}_k,\frac{\sigma_w^2}{LP_p}(\mathbf{D}_{\mathbf{h}_k}^H\mathbf{D}_{\mathbf{h}_k})^{-1}\right),
\end{equation}
which shows that choosing the \ac{RIS} configuration sequences such that $\mx{B}_1$ and $\mx{B}_2$ removes the bias from the \ac{MML} estimator. However, the major setback of this approach is that the minimum number of observations required for this  channel estimation procedure is $2N$ instead of $N$, due to the fact that the $2N$-many $L$-dimensional columns must all be orthogonal to each other, for which $L\geq2N$ is required. Considering that the estimator bias in \eqref{eq:gHat_1} does not vanish with increasing $L$, this is a necessary sacrifice. Hence, it has to be noted that the number of pilot transmissions increases linearly with the number of \acp{RIS} deployed in proximity. In the next section, data signal transmission and its estimation will be analyzed.

\subsection{\ac{MSE} During Channel Estimation}
We consider the \ac{MSE} as the channel estimation performance metric, which is the trace of the error covariance matrix that is derived in this section. In this derivation, we do not assume a particular choice of $\mx{B}_1,\mx{B}_2$, but we utilize the basic assumption $\mx{B}_k\mx{B}_k^H=L\mx{I}_N$. Consequently, we use $\mx{b}_k$ to denote the potential  estimator bias. We can then compute the error covariance matrix as
\begin{align}\nonumber
    \boldsymbol{\Sigma}_{e,k} &= \mathbb{E}\left[(\hat{\mx{g}}_k-\mx{g}_k)(\hat{\mx{g}}_k-\mx{g}_k)^H\right]\\
    &=\mx{b}_k\mx{b}_k^H+\frac{1}{LP_p}\mathbb{E}\left[\mx{D}_{\mx{h}_k}^{-1}\mx{w}_{pk}\mx{w}_{pk}^H\mx{D}_{\mx{h}_k}^{-H}\right]\nonumber\\
    &=\mx{b}_k\mx{b}_k^H+\frac{\sigma_w^2}{LP_p}\left(\mx{D}_{\mx{h}_k}\mx{D}_{\mx{h}_k}^{H} \right)^{-1}.
\end{align}
Consequently, the trace of the error covariance matrix becomes
\begin{equation}\label{eq:traceError}
    \mathrm{tr}(\boldsymbol{\Sigma}_{e,k}) = \|\mx{b}_k\|^2+\frac{\sigma_w^2}{LP_p}\sum_{n=1}^N\frac{1}{|h_{kn}|^2}
\end{equation}
Note that for high $P_p$, $L$, and low $\sigma_w^2$, the second term in \eqref{eq:traceError} vanishes, and the trace of the error covariance converges to $\|\mx{b}_k\|^2$ which depend on the configuration of $\mx{B}_1,\mx{B}_2$:
\begin{equation}\label{eq:chEstFloor}
    \|\mx{b}_k\|^2 = \begin{cases}
        \sum_{n=1}^{N}\frac{|r_{kn}|^2}{|h_{kn}|^2} & \mx{B}_1 = \mx{B}_2,\\
        0 & \mx{B}_1^H\mx{B}_2 = \boldsymbol{0}.
    \end{cases}
\end{equation}
This result shows that configuring the \ac{RIS}s such that $\mx{B}_1^H\mx{B}_2=\boldsymbol{0}$ removes the asymptotic floor on the average \ac{MSE}, which comes from the energy of the estimator bias. On the other hand, when the intended \ac{RIS}-\ac{BS} links $\mx{h}_1,\mx{h}_2$ are strong relative to the unintended and unknown overall link $\mx{r}_k$, the estimator bias will be weaker and the cost of choosing $\mx{B}_1=\mx{B}_2$ will be lower. Nevertheless, pilot contamination results in a fundamental error floor, even if the \acp{RIS} are utilized in different bands. In the next section, we consider the estimation of data based on the channel estimation performed in this section and analyze the consequence of pilot contamination in this phase.

\section{Data Transmission}\label{sec:DataTransmission}

The channel estimation is followed by data transmission. We consider a data packet with a duration shorter than the channel coherence time, therefore, the channels acting on the transmitted data signals are the same as in the channel estimation part. Defining the data signal transmitted by the $k$th \ac{UE} as $x_k\sim\mathcal{CN}(0,1)$, we can express the received data as
\begin{subequations}\label{eq:rxData}
    \begin{align}
        &y_1 = \sqrt{P_d}(\mx{h}_1^T\hat{\boldsymbol{\Phi}}_1\mx{g}_1+\mx{q}_1^T\hat{\boldsymbol{\Phi}}_2\mx{p}_1)x_1 + w_1,\\
        &y_2 = \sqrt{P_d}(\mx{h}_2^T\hat{\boldsymbol{\Phi}}_2\mx{g}_2+\mx{q}_2^T\hat{\boldsymbol{\Phi}}_1\mx{p}_2)x_2 + w_2,
    \end{align}
\end{subequations}
where $w_k\sim\mathcal{CN}(0,\sigma_w^2)$ denotes the receiver noise, $P_d$ denotes the data transmission power, and the RIS configuration matrices $\hat{\boldsymbol{\Phi}}_k$ are selected based on the estimated channels to maximize the average channel gain as shown in \cite[Sec.~II]{Bjornson2022b}:
\begin{align}\nonumber
    &\hat{\phi}_{kn} = \arg(h_{kn}) + \arg(\hat{g}_{kn}),\\
    &\hat{\boldsymbol{\Phi}}_k = \mathrm{diag}\left(e^{-j\hat{\phi}_{k1}},\dots,e^{-j\hat{\phi}_{kN}}\right).\label{eq:RISphasematch}
\end{align}
However, since the \acp{BS} are unaware of the unintended reflections and base their data estimation on channel estimates, they assume the following misspecified received data signal models:
\begin{subequations}\label{eq:rxDataHat}
    \begin{align}
        &\hat{y}_1 = \sqrt{P_d}\mx{h}_1^T\hat{\boldsymbol{\Phi}}_1\hat{\mx{g}}_1x_1 + w_1,\\
        &\hat{y}_2 = \sqrt{P_d}\mx{h}_2^T\hat{\boldsymbol{\Phi}}_2\hat{\mx{g}}_2x_2 + w_2.
    \end{align}
\end{subequations}
Introducing the notation $m_k\triangleq\sqrt{P_d}(\mx{h}_k^T\hat{\boldsymbol{\Phi}}_k\mx{g}_k+\mx{q}_k^T\hat{\boldsymbol{\Phi}}_j\mx{p}_k)$ for $j,k\in\{1,2\}, j\neq k$, and $\hat{m}_k\triangleq\sqrt{P_d}\mx{h}_k^T\hat{\boldsymbol{\Phi}}_k\hat{\mx{g}}_k$, \eqref{eq:rxData} and \eqref{eq:rxDataHat} can be expressed as
\begin{subequations}
    \begin{align}
        &y_k = m_kx_k+w_k,\quad k = 1,2,\label{eq:mData}\\
        &\hat{y}_k = \hat{m}_kx_k+w_k,\quad k = 1,2.\label{eq:mDataHat}
    \end{align}
\end{subequations}
Based on the misspecified observation model in \eqref{eq:mDataHat}, the \acp{BS} estimate $x_k$ by using the misspecified \ac{MMSE} estimator
\begin{equation}
    \hat{x}_k = \frac{\hat{m}_k^*}{|\hat{m}_k|^2+\sigma_w^2}y_k,\quad k = 1,2.
\end{equation}
In this section, we consider the \ac{MSE} between $x_k$ and $\hat{x}_k$ as the performance metric for the data transmission. We derive the data estimation \ac{MSE} for \ac{UE} $k$ as
\begin{align}\nonumber
    &\mathbb{E}\left[|x_k-\hat{x}_k|^2\right] = 1 + \mathbb{E}\left[|\hat{x}_k|^2\right]-2\mathrm{Re}(\mathbb{E}[x_k\hat{x}_k^*])\\
    &=1+\mathbb{E}\left[\frac{|\hat{m}_k|^2(|m_k|^2+\sigma_w^2)}{(|\hat{m}_k|^2+\sigma_w^2)^2}\right]\nonumber\\
    &-2\mathrm{Re}\left(\mathbb{E}\left[\frac{\hat{m}_km_k^*}{|\hat{m}_k|^2+\sigma_w^2}\right]\right)\nonumber\\
    &=1+\mathbb{E}\left[\frac{|\hat{m}_k|^2(|m_k|^2+\sigma_w^2)-2\mathrm{Re}(\hat{m}_1m_1^*)(|\hat{m}_k|^2+\sigma_w^2)}{(|\hat{m}_k|^2+\sigma_w^2)^2}\right]\nonumber\\
    &=1+\mathbb{E}\left[\frac{|\hat{m}_k|^2(|m_k|^2+\sigma_w^2)-2\mathrm{Re}(\hat{m}_1m_1^*)(|\hat{m}_k|^2+\sigma_w^2)}{(|\hat{m}_k|^2+\sigma_w^2)^2}\right]\nonumber\\
    &+\mathbb{E}\left[\frac{\sigma_w^2(|m_k|^2+\sigma_w^2)-\sigma_w^2(|m_k|^2+\sigma_w^2)}{(|\hat{m}_k|^2+\sigma_w^2)^2}\right]\nonumber\\
    &=1+\mathbb{E}\left[\frac{(|\hat{m}_k|^2+\sigma_w^2)(|m_k|^2+\sigma_w^2-2\mathrm{Re}(\hat{m}_km_k^*))}{(\hat{m}_k^2+\sigma_w^2)^2}\right]\nonumber\\
    &-\mathbb{E}\left[\frac{\sigma_w^2(|m_k|^2+\sigma_w^2)}{(\hat{m}_k^2+\sigma_w^2)^2}\right]\nonumber\\
    &=\mathbb{E}\left[\frac{|m_k-\hat{m}_k|^2+2\sigma_w^2}{|\hat{m}_k|^2+\sigma_w^2}\right]-\mathbb{E}\left[\frac{\sigma_w^2(|m_k|^2+\sigma_w^2)}{(|\hat{m}_k|^2+\sigma_w^2)^2}\right].\label{eq:dataMSE}
\end{align}
Defining $\epsilon_k\triangleq m_k-\hat{m}_k$, \eqref{eq:dataMSE} can be rewritten as
\begin{equation}\label{eq:dataMSE_v2}
    \mathbb{E}[|x_k-\hat{x}_k|^2] = \mathbb{E}\left[\frac{|\epsilon_k|^2+2\sigma_w^2}{|m_k-\epsilon_k|^2+\sigma_w^2}-\frac{\sigma_w^2(|m_k|^2+\sigma_w^2)}{(|m_k-\epsilon_k|^2+\sigma_w^2)^2}\right]
\end{equation}
To examine the impact of pilot contamination on the data estimation performance more clearly, we now consider  channel estimation at high \acp{SNR}, so that the estimation error only comes from the estimator bias, i.e., pilot contamination. This happens  when $L$ or $P_p$ is high and/or $\sigma_w^2$ is low, which results in that the estimator covariances in \eqref{eq:gHat_1} and \eqref{eq:gHat_2} become zero.  For notational convenience, we consider the case where $P_p$ is arbitrarily large so that  $\lim_{P_p\to\infty}\hat{\mx{g}}_k = \mx{g}_k+\mx{b}_k$, where
\begin{equation}\label{eq:biasCases}
    \mx{b}_k = \begin{cases}
        \mx{D}_{\mx{h}_k}^{-1}\mx{D}_{\mx{q}_k}\mx{p}_k & \mx{B}_1=\mx{B}_2,\\
        \boldsymbol{0} & \mx{B}_1^H\mx{B}_2=\boldsymbol{0}.
    \end{cases}
\end{equation}
\subsection{Data \ac{MSE} with Channel Estimation at High \ac{SNR}}
In \eqref{eq:dataMSE_v2}, $\epsilon_k$ and $m_k$ are functions of $\hat{\mx{g}}_1$ and $\hat{\mx{g}}_2$, therefore as $\hat{\mx{g}}_1$ and $\hat{\mx{g}}_2$ converge to their means, $\epsilon_k$ and $m_k$ become
\begin{subequations}\label{eq:barchannels}
\begin{align}
    &\overline{m}_k = \sqrt{P_d}(\mx{h}_k^T\Bar{\boldsymbol{\Phi}}_k\mx{g}_k+\mx{q}_k^T\Bar{\boldsymbol{\Phi}}_j\mx{p}_k),\\
    &\overline{\epsilon}_k = \sqrt{P_d}(\mx{q}_k^T\Bar{\boldsymbol{\Phi}}_j\mx{p}_k-\mx{h}_k^T\Bar{\boldsymbol{\Phi}}_k\mx{b}_k),
\end{align}    
\end{subequations}
for $j,k\in\{1,2\}$ and $j\neq k$. $\Bar{\boldsymbol{\Phi}}_k$ denotes the \ac{RIS} configuration computed according to \eqref{eq:RISphasematch} with $\hat{\mx{g}}_k=\mx{g}_k+\mx{b}_k$. \newpage At high \ac{SNR}, the MSE in \eqref{eq:dataMSE_v2} can be rewritten as
\begin{equation}\label{eq:dataMSE_highChan}
    \text{MSE} = \frac{|\overline{\epsilon}_k|^2+2\sigma_w^2}{|\overline{m}_k-\overline{\epsilon}_k|^2+\sigma_w^2}-\frac{\sigma_w^2(|\overline{m}_k|^2+\sigma_w^2)}{(|\overline{m}_k-\overline{\epsilon}_k|^2+\sigma_w^2)^2}.
\end{equation}
This is a practically achievable limit since RIS-aided systems require large pilot sequences over a narrow bandwidth, so the SNR might be larger than in the data transmission phase.

\subsection{Data \ac{MSE} with Transmission at 
High \ac{SNR}}
In the previous subsection, we obtained the expression for data \ac{MSE} when  the channels are estimated at a high pilot \ac{SNR}, while the data transmission is done at an arbitrary SNR. To study the case when also the data transmission a step further, we let $\sigma_w^2 \to 0$, which results in the limit
\begin{equation}\label{eq:highSNRfloorData}
    \lim_{\sigma_w^2\to 0}\text{MSE} = \frac{|\overline{\epsilon}_k|^2}{|\overline{m}_k-\overline{\epsilon}_k|^2}.
\end{equation}
Note that the resulting expression denotes the ratio between the estimated overall \ac{SISO} channel $\hat{m}_k$'s power and the mismatch parameter $\epsilon_k$'s power. Recall that $\mx{b}_k$ depends on which \ac{RIS} pilot sequence is utilized. For $\mx{B}_1=\mx{B}_2$, we can obtain $\epsilon_k$ as
\begin{align}\nonumber
    \epsilon_k &= \sqrt{P_d}\mx{q}_k^T\hat{\boldsymbol{\Phi}}_j\mx{p}_k-\sqrt{P_d}\mx{h}_k^T\hat{\boldsymbol{\Phi}}_k\mx{D}_{\mx{h}_1}^{-1}\mx{D}_{\mx{q}_1}\mx{p}_1\nonumber\\
    &=\sqrt{P_d}\mx{q}_k^T\hat{\boldsymbol{\Phi}}_j\mx{p}_k-\sqrt{P_d}\hat{\boldsymbol{\phi}}_k\mx{D}_{\mx{h}_k}\mx{D}_{\mx{h}_k}^{-1}\mx{D}_{\mx{q}_1}\mx{p}_1\nonumber\\
    &=\sqrt{P_d}\mx{q}_k^T\hat{\boldsymbol{\Phi}}_j\mx{p}_k-\sqrt{P_d}\hat{\boldsymbol{\phi}}_k\mx{D}_{\mx{q}_1}\mx{p}_1\nonumber\\
    &=\sqrt{P_d}\mx{q}_k^T\hat{\boldsymbol{\Phi}}_j\mx{p}_k-\sqrt{P_d}\mx{q}_k^T\hat{\boldsymbol{\Phi}}_k\mx{p}_k\nonumber\\
    &=\sqrt{P_d}\mx{q}_k^T(\hat{\boldsymbol{\Phi}}_j-\hat{\boldsymbol{\Phi}}_k)\mx{p}_k.
\end{align}
On the other hand, $\mx{B}_1^H\mx{B}_2=0$ removes $\mx{b}_k$ for $k\in\{1,2\}$. Consequently, we have
\begin{equation}\label{eq:epsilonExplicit}
    \epsilon_k = \begin{cases}
        \sqrt{P_d}\mx{q}_k^T(\hat{\boldsymbol{\Phi}}_j-\hat{\boldsymbol{\Phi}}_k)\mx{p}_k & \mx{B}_1=\mx{B}_2,\\
        \sqrt{P_d}\mx{q}_k^T\hat{\boldsymbol{\Phi}}_j\mx{p}_k & \mx{B}_1^H\mx{B}_2=\boldsymbol{0}.
    \end{cases}
\end{equation}
It is possible to observe that the $\epsilon_k$ corresponds to only the unintended reflection path itself when $\mx{B}_1^H\mx{B}_2=\boldsymbol{0}$. On the other hand, $\mx{B}_1=\mx{B}_2$ yields an expression depending on the difference between the two \acp{RIS}' configurations during data transmission. Since each \ac{RIS} is configured based on the channels of their respective users, it is highly unlikely that the configurations will be close. Moreover, it has to be noted that the \ac{RIS} configuration of the non-serving \ac{RIS} is different among the two cases since the channel estimates are also different.

\section{Numerical Results}\label{sec:NumRes}
In this section, numerical results are provided to demonstrate the impact of pilot contamination in both the channel and data estimation phases. We consider $N=256$ \ac{RIS} elements. First, we demonstrate the results for the trace of the channel estimation error covariance matrix (i.e., the sum \ac{MSE}). Then, for a single set of channel realizations, we provide the data estimation \ac{MSE} for different data transmission powers. We also provide a \ac{CDF} plot for the high-\ac{SNR} data estimation \ac{MSE} floors under $\mx{B}_1=\mx{B}_2$ and $\mx{B}_1^H\mx{B}_2=\boldsymbol{0}$. The list of parameters used can be found in Table \ref{tab:chanEstParams}.

\subsection{Channel Estimation}\label{sec:NumResChanEst}
\begin{figure}
    \centering
    \includegraphics[width=0.84\linewidth]{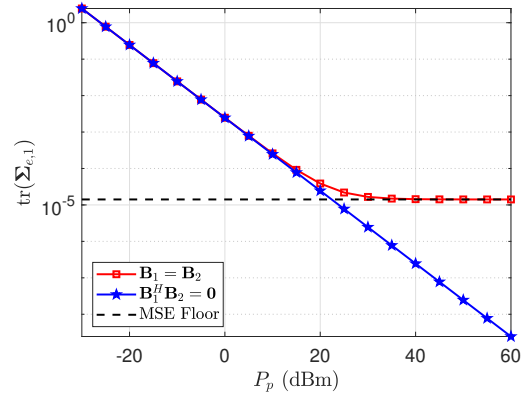}
    \caption{Pilot transmission power versus channel estimation \ac{MSE}.}
    \label{fig:chanEstSNR}
\end{figure}

For the channel estimation, we consider the \ac{MSE} as our performance metric. Moreover, we consider the results for a single \ac{UE}, since the results for different \acp{UE} only differ by the channel realizations. 
\begin{table}[t]
    \centering
    \caption{Parameters used in the numerical results.}
    \begin{tabular}{|c|c|}
        \hline
        Parameter & Value\\
        \hline
        $P_p$ or $P_d$ & $-30,-25,\dots,40$ dBm\footnote{The results for $P_p = 45,50,55,\text{ and }60$ dBm are also demonstrated in Fig. \ref{fig:chanEstSNR} to display the high \ac{SNR} floor more clearly.}\\
        \hline
        \ac{UE}-\ac{RIS} path loss & $-80$ dB\\
        \hline
        \ac{RIS}-\ac{BS} path loss & $-60$ dB\\
        \hline
        $\sigma_w^2$ & $-90$ dBm\\
        \hline
        $N$ & $256$\\
        \hline
        $L$ & $513$\\
        \hline
    \end{tabular}
    \label{tab:chanEstParams}
\end{table}
In Fig. \ref{fig:chanEstSNR}, we plot \eqref{eq:traceError} for different values of $P_p$, and we also provide the high-\ac{SNR} floor for the case where $\mx{B}_1=\mx{B}_2$. Note that at lower transmission powers, the covariance matrix of the estimator acts dominantly, hence, both \ac{RIS} configurations perform nearly the same. However, after $P_p=20$ dBm, the power of the estimator bias starts to dominate, and the average \ac{MSE} for $\mx{B}_1=\mx{B}_2$ goes to the floor denoted by the black dashed line, which is given by \eqref{eq:chEstFloor}. On the other hand, the average \ac{MSE} for $\mx{B}_1^H\mx{B}_2=\boldsymbol{0}$ does not stop there but keeps decreasing towards zero. As mentioned before, the \ac{MML} estimators used by the \acp{BS} coincide with the true \ac{ML} estimators when the \acp{RIS} are configured such that $\mx{B}_1^H\mx{B}_2=\boldsymbol{0}$.
\subsection{Data Estimation}\label{sec:NumResDataEst}
\begin{figure}
    \centering
    \includegraphics[width=0.84\linewidth]{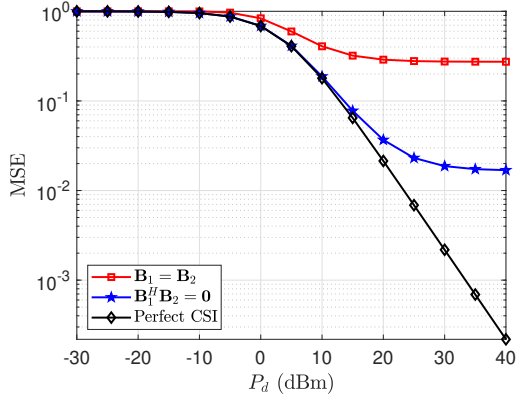}
    \caption{Data transmission power versus data estimation \ac{MSE} with high channel estimation \ac{SNR}.}
    \label{fig:dataMSESNR}
\end{figure}

In Fig. \ref{fig:dataMSESNR}, the data estimation \ac{MSE} performance with the two \ac{RIS} pilot configurations are analyzed when the channel estimation \ac{SNR} is high. That is, \eqref{eq:dataMSE_highChan} is plotted for $\mx{B}_1=\mx{B}_2$ and $\mx{B}_1^H\mx{B}_2=\boldsymbol{0}$. In addition, the case where all of the channels are perfectly known is plotted to serve as the golden standard, labeled as \emph{Perfect \ac{CSI}}. However, even when all the channels are perfectly known, each \ac{RIS} is assumed to be optimized according to the subscribed \ac{UE}'s \ac{CSI}. Note that although the channel estimation \ac{SNR} is high, $\mx{B}_1=\mx{B}_2$ yields biased estimates of $\mx{g}_1$ due to pilot contamination caused by self-interference. On the other hand, $\mx{B}_1^H\mx{B}_2=\boldsymbol{0}$ yields the true $\mx{g}_1$ as the estimate, however, since \ac{BS} $1$ is unaware of the path through the second \ac{RIS}, the data estimate is biased, hence, there is still a high data transmission \ac{SNR} floor. At around $P_d=5$ dBm, $\mx{B}_1=\mx{B}_2$ starts to approach the high-\ac{SNR} floor. On the other  hand, $\mx{B}_1^H\mx{B}_2=\boldsymbol{0}$ does not suffer from the lack of awareness of the second \ac{RIS} path until around $P_d=20$ dBm. Hence, Fig. \ref{fig:dataMSESNR} clearly shows the benefit of configuring the \ac{RIS} pilot configurations sequences orthogonally. 

\begin{figure}
    \centering
    \includegraphics[width=0.84\linewidth]{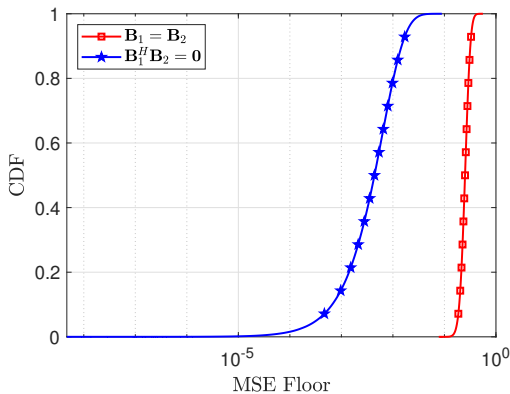}
    \caption{\ac{CDF} of the data-\ac{MSE} floors for two \ac{RIS} pilot configurations for i.i.d. Rayleigh fading.}
    \label{fig:highSNRCDF}
\end{figure}

Note that \eqref{eq:highSNRfloorData} and \eqref{eq:epsilonExplicit} do not guarantee the superiority of $\mx{B}_1^H\mx{B}_2=\boldsymbol{0}$ over $\mx{B}_1=\mx{B}_2$, since if both \acp{RIS} were configured identically during the data transmission phase, $\mx{B}_1=\mx{B}_2$ would not suffer from a high-\ac{SNR} data estimation \ac{MSE} floor. To demonstrate that this scenario is highly unlikely, empirical \acp{CDF} of the \ac{MSE} floors at high \ac{SNR} are provided. We generate each channel according to $\mathcal{CN}(\boldsymbol{0},\mx{I}_N)$, and then scale them according to the path losses given in Table \ref{tab:chanEstParams}. The resulting \acp{CDF} are provided in Fig. \ref{fig:highSNRCDF}. 
This figure is generated by using $10^6$ different sets of channel realizations for $N=32$ \ac{RIS} elements. With a probability less then $10^{-6}$, the high-\ac{SNR} floor under $\mx{B}_1^H\mx{B}_2=\boldsymbol{0}$ is much lower than that of $\mx{B}_1=\mx{B}_2$, clearly demonstrating the benefit of using orthogonal \ac{RIS} pilot configurations over identical configurations.
\section{Conclusions}\label{sec:Conclusion}
In this paper, we have studied the impact of pilot contamination in a system consisting of two wide-band \acp{RIS}, two single-antenna \acp{UE}, and two co-located single-antenna \acp{BS}. We have demonstrated that the presence of multiple \acp{RIS} in the same area causes pilot contamination, although the \acp{UE} are subscribed to different operators and transmit over disjoint narrow frequency bands. To combat this new type of pilot contamination, we proposed the use of orthogonal \ac{RIS} configurations during pilot transmission. In the numerical results, we have clearly shown that the proposed approach eliminates pilot contamination completely, and decreases data estimation \ac{MSE} significantly. While increasing the number of pilots to configure \acp{RIS} orthogonally alleviates pilot contamination, more efficient ways of dealing with this problem are needed in the future.
\bibliographystyle{IEEEtran}
\bibliography{IEEEabrv,main}
\end{document}